\documentclass[]{article}

\usepackage{authblk}
\usepackage{graphicx}
\usepackage{epstopdf}% To incorporate .eps illustrations using PDFLaTeX, etc.
\usepackage[caption=false]{subfig}% Support for small, `sub' figures and tables
\begin{document}

%\articletype{ARTICLE TEMPLATE}% Specify the article type or omit as appropriate

\title{Combining Audio and Non-Audio Inputs in Evolved Neural Networks for Ovenbird Classification \footnote{This is a preprint version of the article published in Bioacoustics Volume 33 Issue 3, April 9 2024, available online: https://www.tandfonline.com/doi/pdf/10.1080/09524622.2024.2329193}}

%\author{
%\name{Sergio Poo Hernandez\textsuperscript{a}\thanks{CONTACT Sergio Poo Hernandez. Email: pooherna@ualberta.ca}, Vadim Bulitko\textsuperscript{a} and Erin Bayne\textsuperscript{b}}
%\affil{\textsuperscript{a}Department of Computing Science, University of Alberta, Edmonton, Canada; %\textsuperscript{b}Department of Biological Sciences, Univertsity of Alberta, Edmonton, Canada}
%\date{}
%}

\author[1]{Sergio Poo Hernandez}
\author[2]{Vadim Bulitko}
\author[3]{Erin Bayne}
\affil[1]{Department of Computing Science, University of Alberta, Edmonton, Canada}
\affil[2]{Department of Computing Science, University of Alberta, Edmonton, Canada}
\affil[3]{Department of Biological Sciences, University of Alberta, Edmonton, Canada}
\date{}

\maketitle

\begin{abstract}
In the last several years the use of neural networks as tools to automate species classification from digital data has increased. This has been due in part to the high classification accuracy of image classification through Convolutional Neural Networks (CNN). In the case of audio data CNN based recognizers are used to automate the classification of species in audio recordings by using information from sound visualization (i.e., spectrograms). It is common for these recognizers to use the spectrogram as their sole input. However, researchers have other non-audio data, such as habitat preferences of a species, phenology, and range information, available that could improve species classification. In this paper we present how a single-species recognizer neural network’s accuracy can be improved by using non-audio data as inputs in addition to spectrogram information. We also analyze if the improvements are merely a result of having a neural network with a higher number of parameters instead of combining the two inputs. We find that networks that use the two different inputs have a higher classification accuracy than networks of similar size that use only one of the inputs.only one of the inputs.
\end{abstract}

%\begin{keywords}
%Sections; Birdsong; classification; machine learning; spectrogram
%\end{keywords}

\section{Introduction}

For decades human observers have gone to the field to count birds.  Typically this is done using point counts where individuals of each species seen and heard during a fixed period of time are counted.  The limited time spent in any place means many individuals and even species are not detected by observers even though they are present.  A large effort has gone into developing statistical techniques to correct for the resulting detection error.  However, the assumptions of such statistical corrections can be difficult to meet.  Consequently ornithologists are trying to find ways to minimize detection error by improving how they count birds.

The majority of birds counted by human observers come from aural cues.   Autonomous recording units (ARUs) are a possible solution to the detection error problem for animals that make sounds. With increasing recording effort and processing of audio files one can be increasingly confident that all individuals and species get counted.  Listening to recordings multiple times improves the accuracy of identification and improves our ability to count individuals.  However, given the large amount of recordings coming from ARUs, it creates the challenge of processing the data in a cost-effective manner.  Consequently there has been an increasing effort to use computers to help find and identify bird sounds in digital audio recordings, a process known as automated classification. Various approaches to automated classification of animal sounds have been developed including Hidden Markov models~\cite{chu2011noise}, k-nearest neighbors~\cite{huang2009frog,bang2014classification}, support vector machines~\cite{acevedo2009automated,armitage2010comparison} and random forests~\cite{armitage2010comparison,noda2016automatic}. Non-machine learning approaches including band-limited energy detection~\cite{charif2010raven}. Recognizers use automated classifiers to go over recordings to identify when a bird sings. This data is then used to determine presence of a bird in the recording and therefore in the site of the recording.

Most recognizers rely on spectrograms as input and increase performance by focusing on improving the training data through spectrogram manipulation. Some recognizers utilize Mel Frequency Cepstral Coefficients~\cite{cheng2010call}, time and frequency domain features such as signal bandwidth or spectral centroid~\cite{huang2009frog}, time-frequency features such as signal minimum and maximum frequency~\cite{acevedo2009automated}, or specialized approaches such as ridge-based features~\cite{dong2015similarity,xie2015application}.  What has not been done is to automatically use the additional non-audio information that ornithologists rely on in the field to determine which species they are hearing.  Decades of point counts have taught ornithologists how likely they are to hear birds of a particular species at a certain time of year, time of day, in a particular spatial location and how vegetation conditions influence detection error and abundance. Whether giving such data to the network is beneficial to better classification is an open research question.  We conjecture that training neural networks with this information in addition to the spectrogram can improve the ability of recognizers to correctly find detections of specific species in digital audio recordings. 

To answer this question we make the following contributions. First,  we developed a neural network to recognize Ovenbirds (\textit{Seiurus aurocapilla}) songs in audio files.  Specifically, we first developed a network that learned solely based on spectrograms of Ovenbird songs. We also trained a network that used environmental information such as time of year the recording was made, vegetation conditions that influence whether Ovenbirds are likely to be present, areas in and outside of Ovenbird range, and recordings at times of day when Ovenbirds do not typically call. We finally trained a network that combined both types of information, the spectrogram and environmental information. 

We then analyzed how classification accuracy of the networks changed as the size of the networks increased. The main question we address in this paper is, given limitations on how big the network can be, is it worth it to allocate parameters in a network to use data other than the spectrogram input to improve the classification accuracy of the recognizing classifier or is it better to use all parameters to get better classification accuracy with higher resolution spectrograms?

\section{Our Approach}

We present our approach in two parts. First, we discuss  how we process the data. Second, we discuss the selection of the network architecture. We use simulated evolution to search a space of possible architectures to find the best performing ones. 

\subsection{Data processing}
Commonly neural networks used as recognizers in sound identification tasks have spectrograms as their input~\cite{knight2020pre}. The main differences among these recognizers are how the input is handled and the network’s architecture. These differences were introduced to improve the quality of the input signal by removing noise to isolate the clean signal, as well as training the network with noisy data to improve accuracy~\cite{kahl2021birdnet}. A common practice for selecting network architectures is to modify neural networks that have been used on similar tasks~\cite{kahl2021birdnet,schwab2022automated}. However, there is information about the environment in the recordings that could potentially help make a more accurate recognizer. Some of this information is easily obtained from the recordings themselves such as the time of day and day of the year as well as the coordinates of recorder location. The coordinates can then be used to obtain detailed habitat information such as the type of vegetation, climate, and elevation.

\begin{figure}
\centering
\resizebox*{14cm}{!}{\includegraphics{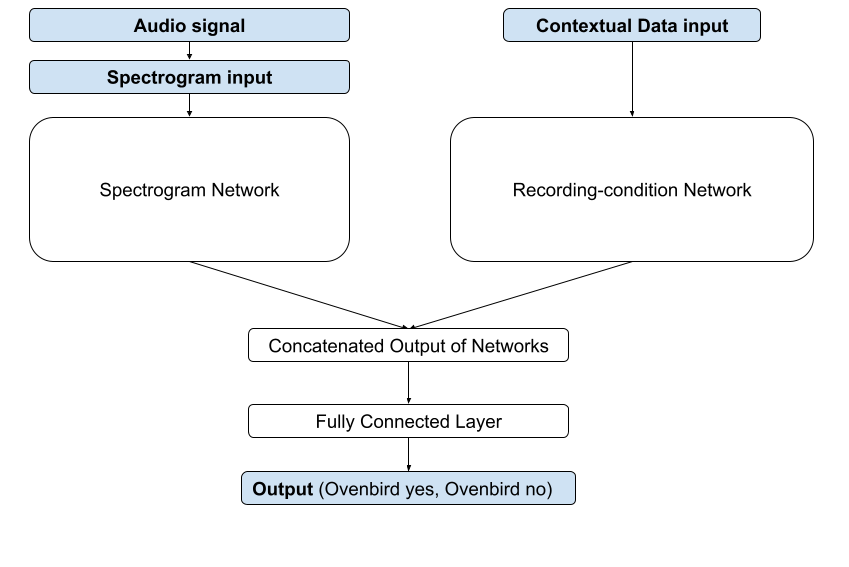}}\hspace{5pt}
\caption{} \label{fig:Networks}
\end{figure}

Consequently our approach uses two inputs: 1) the spectrogram image of the audio signal and 2) information about where and when the recording was obtained. To handle the two different inputs the network’s architecture consists of two sets of layers, referred to as columns: one column for analyzing the spectrogram input and the other for the other information about the recording. Then the two outputs of the columns are concatenated and used to make the final classification (Figure~\ref{fig:Networks}).

An expert went through the recordings to identify and tag Ovenbird songs from which we created a training data set. Using the information from the tag  we cut out from the recording the clips that were used for training. For the files in the dataset that contain Ovenbird calls we ensured that the full call happens somewhere in the clip. For the files in the dataset that do not contain Ovenbird calls we took a random clip from recordings where no Ovenbirds were ever detected as well as random clip before Ovenbird songs for recordings where Ovenbirds were known to be present and vocalizing. We clip before the first Ovenbird song because for some of the recordings we could only guarantee there were no Ovenbird songs before the first tag.

\begin{figure}
\centering
\resizebox*{14cm}{!}{\includegraphics{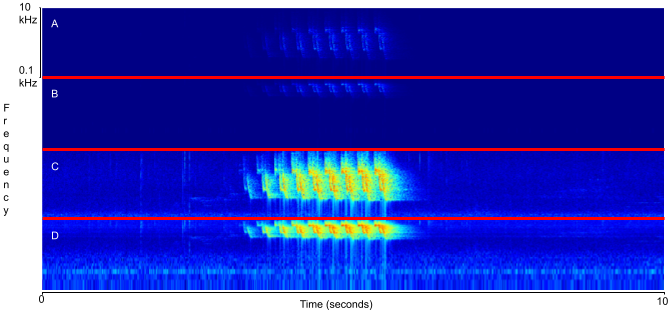}}\hspace{5pt}
\caption{} \label{fig:spect}
\end{figure}

For the spectrogram input we adopted the approach from our previous work~\cite{knight2020pre} in which we created four different spectrograms per audio recording and put them into a single image file. The difference among the spectrograms was the application of log scales to the frequency and amplitude of the signal. However, instead of arranging them in quadrants as before, we stacked them vertically (Figure~\ref{fig:spect}). This change made it easier to maintain the dataset since it makes trimming an audio recording easier.

For the recording conditions we used the times of recordings as well as habitat data extracted from a GIS layer~\cite{abmi}. This data was stored in a table listing all of the recording conditions at locations and times of the recordings. The table contained 41 variables per recording and includes information on vegetation type, time since last disturbance, type and amount of human disturbance. Time of day and day of year were also included. The information was both numerical and categorical.

We ensured the data from the two categories (no Ovenbird and Ovenbird observed) were balanced in terms of the number of clips. This was done to prevent the network from learning to classify by choosing the class with the higher number of samples. We also made sure to include clips that lacked Overbird songs but had the same recording conditions as the clips in which there were Ovenbird songs.  This was done so that the network did not learn to discriminate based solely on the recording conditions.

\subsection{Evolution of Neural Network Architectures}
We were also interested in evaluating how network architecture size affects classification accuracy. In order to investigate this systematically we needed to have several networks of different sizes. To have a broad range of architectures we chose to automatically generate network architectures for both types of inputs using a simulated evolution.

Specifically, simulated evolution can be used to search the space of architectures for each column to trade network performance for network size. The process starts by creating the first generation of networks using randomly generated architectures. Then they were evaluated to determine their fitness which is a linear combination of classification accuracy (higher is better) and network size as the number of learnable parameters (smaller is better).  Fittest performing networks were selected for the next generation and the rest were discarded. Each successive generation was populated by the fittest individuals surviving from the previous generation and their offspring obtained by applying random mutations to the survivors. The mutations took the form of removing, adding and/or modifying the layers in the architecture. The process was repeated for a predetermined number of generations or stopped early if the classification accuracy of the top performer across successive generations stalled..

Spectrogram networks were composed of convolutional layers and/or fully-connected layers while the recording-condition networks were composed of only fully-connected layers. Evolved spectrogram architectures had a general architecture of convolutional layers, if used since a network without convolutional layers could be evolved, followed by one or more fully-connected layers. We evolved architectures of each column individually looking for the fittest architecture in classifying Ovenbird songs based on the spectrogram of the audio recording and another one based on the recording conditions-only data. We then used these two best performing networks by themselves as well as concatenated them to form a combined network (Figure~\ref{fig:Networks}).

\section{Empirical Evaluation}

Audio clips were obtained from recordings collected through ARUs by members of the Bioacoustic Unit~\cite{wildtrax}. Our data consisted of 2769 audio recordings. For each recording we created spectrograms from 10 second clips where an Ovenbird was known to be vocalizing and known not to be present and/or vocalizing. The spectrograms had a size of 224 by 224 pixels in 3 color channels. Our dataset totaled 26362 spectrograms from the 2769 audio recordings (12866 that contained an Ovenbird song and 13496 that contained no Ovenbird songs). The number of spectrograms was higher than recordings because a single recording could contain multiple Ovenbird songs and a spectrogram was created for each song tagged in the recording. As for the spectrograms that did not contain Ovenbird songs we created spectrograms from clips so the number of spectrograms would match those with Ovenbird songs. Of the spectrograms that did not have Ovenbird songs, 6360 were from recordings that have Ovenbird songs elsewhere in them but not in the 10 second clip from which the spectrogram was generated.

We evolved several network architectures for spectrogram and recording-condition networks. We ran the evolution three times with the following three fitness functions. The first fitness function included classification accuracy only. The second fitness function was network size only. The third fitness function was a linear combination of the two.  The objective was to obtain a variety of networks with different classification accuracy and sizes in order to establish relation between network size and classification accuracy. Each evolved network is trained using 75\% of the clips, the other 25\% is used for testing the classification accuracy. The training data is further split into training data (75\%) and validation data (25\%). The validation data is used to avoid overfitting of the network during training. We used the Adam optimizer for training with a batch size of 20 samples for up to 50 epochs.

Table~\ref{tab:accuSpect} lists the classification accuracy for a low, medium and high accuracy spectrogram-only networks. The architectures of the low and medium accuracy networks were evolved while the architecture of the high accuracy network was a modified DenseNet-47~\cite{huang2017densely}. The modifications to DenseNet-47 were to the input and output layers so the network would work with our spectrograms and provide ovenbird/no ovenbird output.

We also evolved network architecture that uses the recording-condition data as the input instead of a spectrogram. The lowest accuracy recording-condition architecture has about the same accuracy (51.3\%)  as the lowest-accuracy spectrogram-only network with 50.1\% accuracy. However, the evolved recording-condition only networks outperform the spectrogram networks of similar size when the networks have fewer than a million parameters. For larger networks the spectrogram networks outperform recording-condition only networks of similar size as the recording conditions-only networks’ classification accuracy plateau at 61.7\%. 

\begin{figure}
\centering
\resizebox*{14cm}{!}{\includegraphics{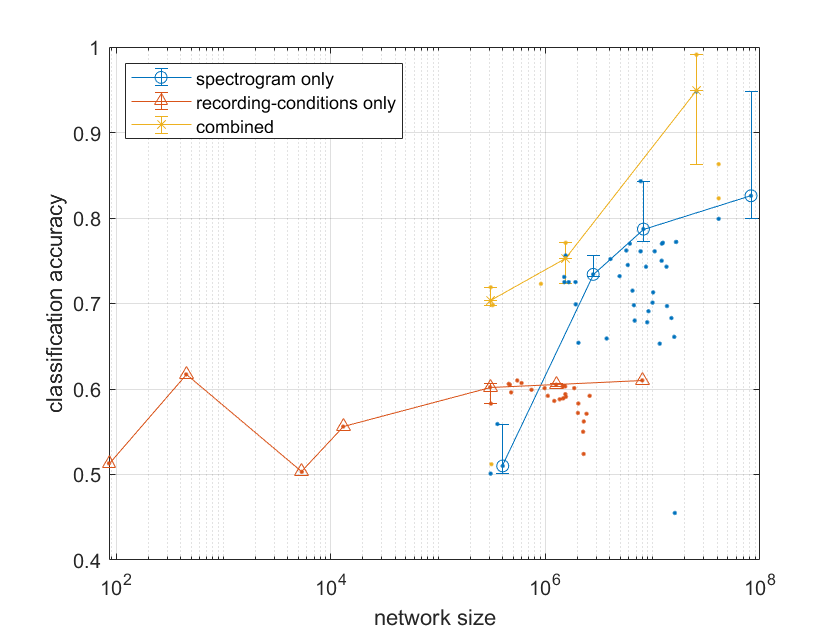}}\hspace{5pt}
\caption{} \label{fig:result}
\end{figure}

Classification accuracy of the different networks is shown in Figure~\ref{fig:result} which plots classification accuracy as  a function of network size. The markers show the median performer out of the top-three networks in categories based on the network size. The error bars at each marker show the difference between the median value and the top performing network and the third best performing network in the category. This illustrates the difference in the accuracy of the top three networks in each category. The blue circle markers show the results for the spectrogram-only networks, the red triangles are for the recording-condition only networks and yellow stars are for the combined networks. The spectrogram-only network classification accuracy ranges from 50\% to 94.8\%. There is a trend of larger networks delivering a higher classification accuracy. For the recording-condition only networks the accuracy ranges from 51\% to 61.7\%. 

Finally, Figure~\ref{fig:result} suggests that combined networks can outperform the networks that use either input type alone, for similar network size (networks with similar number of trainable parameters). The lines connecting the markers are linear interpolations and as shown in the figure there are single input networks that perform better than the combined networks, and there could be more that were not encountered during evolution. However, given that most of the network architectures we did find have performances below the interpolated line we have evidence that dedicating some of the network’s parameters to process data other than the spectrogram, the recording conditions data in our case, can improve the network’s classification accuracy.

\section{Discussion}

\subsection{Recording-condition only networks}
Given the results for the recording condition-only  networks in Table~\ref{tab:accuCond} it seems the network is capable of determining which habitats do not have Ovenbirds in them, however it struggles with habitats where Ovenbirds might sing but did not happen to vocalize during the recording. This suggests that it might be difficult or impossible to determine whether an Ovenbird is present and vocalizing by using only the  recording-condition information. To investigate this point we looked at the true negatives and false negatives given by the network. Table~\ref{tab:confCond} and Table~\ref{tab:confCondBD} suggest that a network is able to learn to recognize locations where Ovenbirds never sing quite easily. However, given the  recording context information for locations where Ovenbirds can sometimes sing, the network is unable to reliably determine whether or not an Ovenbird is present and vocalizing in the section of recording. This does not occur with the locations where Ovenbirds were never observed to sing as the network classifies most recordings as having no Ovenbird songs, since it learns that those conditions will not have Ovenbirds present. In other words, the networks are able to learn in which locations Ovenbirds will never be present. 

\subsection{Spectrogram-only and combined networks}
Table~\ref{tab:confSpec} and~\ref{tab:confSpecBD} show a similar breakdown of the categories for the best performing spectrogram-only network. The results show that the network overall classifies spectrograms correctly. However, it does generate a higher number of false positives than false negatives. False positives occur more frequently for the non-Ovenbird habitat spectrograms (Table~\ref{tab:confSpecBD}).

A similar analysis of the combined-input network indicates that adding the recording context as an input allows a network to outperform the spectrogram-only networks. There is a considerable increase in the true negatives from the combined network (Table~\ref{tab:confCombBD}). The results seem to indicate that the combination of inputs helps the network learn that there are habitats where Ovenbirds do not sing and therefore make it less likely to confuse a non-Ovenbird audio clip with an Ovenbird audio clip.

\section{Future Work}

In wildlife management biologists typically use transcribed data from ARUs to create a metric of occurrence and abundance of a species that is then used as the dependent variables in regression style analysis using environmental data as the response variables.  Increasingly biologists are using automated recognizers to get an occurrence/ abundance index. The study presented in this paper suggests that inputting non-audio data into a neural net may increase recognition accuracy.

A potential concern of this approach is that the inclusion of environmental variables in the recognition task may influence the predictions from regression-based models of species - environment relationships when the recognizer is used to estimate occurrence and abundance.  For example, our recognizer might suggest an Ovenbird is not present in a location primarily because of the environmental inputs while the spectrogram only recognizer might show the species was present in an unexpected environmental condition. This could in turn  result in regression models based on occurrence/ abundance outputs from the recognizer having zeros in vegetation types that actually did have the species of interest.  This could create a degree of circularity where what the neural network thinks is important as bird habitat drives what the regression-based model learns is important habitat This emphasizes how the range of environmental conditions the neural network is presented with in training will matter.  Different vegetation types might be used by a species outside where the neural network was trained and could exclude detections based on the fact that the vegetation conditions are unsuitable in the area where the recognizer was trained.  Similarly, if a species changed vegetation types over time in the area where a recognizer was built, signals in previously unused vegetation types could be missed as the animal alters what it selects for vegetation over time.   However, currently used recognizers based solely on spectrogram inputs require human validation to be highly accurate.  We propose that validation of results from the recognizer still is required in most applications.  Identifying false positives found by the recognizer and listening to a subset of the recordings to find those false positives can provide information to improve the recognizer by expanding the environmental conditions where species occur.

To assess how much the environmental variables influence outcomes we should test how well a trained network recognizes Overnbirds in a novel environment, unseen during its training. It is possible that the classification accuracy would degrade substantially, requiring a re-training, particularly if vegetation conditions in the new environment are substantially different from those in the training data.  In addition, birds do show variation in habitat selection, using different vegetation types in different parts of their range which could require recognizers that account for spatial covariates like latitude and longitude. However adding this new information may require the retraining of the network, due to a phenomenon known as catastrophic forgetting~\cite{kirkpatrick2017overcoming}.

Presently whenever we add an input to a network we first search for a new network architecture from scratch using simulated evolution. We then train the evolved architecture from scratch on the new training data. As a result both previously evolved network architecture and all of its learned weights are discarded. Using ideas by~\cite{kirkpatrick2017overcoming} future work will investigate preserving evolved architecture and learned weights and extending them with new parts of the architecture and new weights, all without having to re-evolve and re-train everything from scratch.

Another direction for future work is to obtain recording-condition data automatically (e.g., from satellite imagery for the recording location).  Using satellite imagery directly to aid in species classification may result in even larger networks making incremental evolution and training even more important.  Such imagery would use considerably more parameters than the 41 variables we used and relative trade-offs in spectrogram vs remote-sensing image resolution might become more apparent

\section{Conclusion}
Our empirical evaluation suggests that adding the recording-condition information as an input can improve classification accuracy over spectrogram-only networks, controlling for the network size. Specifically the results suggest that for a small network size recording-condition only networks outperform audio-only networks. For a larger network size audio-only networks are better than recording-condition only networks yet networks that use both inputs are more accurate still. So the recipe for a practitioner may be to use only a recording-condition input when the network must be small in size but use both recording-condition and audio inputs for larger sizes.

\section*{Acknowledgements}
We appreciate support by NSERC and Nvidia.

\bibliographystyle{plain}
\bibliography{ovenbird}

\appendix
\section{Neural Network Architecture Evolution Details}

The spectrogram-only networks have a lower bound for the number of parameters the network can have. This is because of the parameters used to setup convolution filters that are used by the evolution process to build the network, the filters are padded with zeros to fit the image. Due to these constraints networks with fewer parameters are not part of the space of possible architectures evolution searches in. The evolution space is defined by the following parameters:
\begin{itemize}
\item \textbf{Maximum number of convolutional layers:} 4
\item \textbf{Maximum number of fully connected layers:} 4
\item \textbf{Possible dropout bales:} 0, 0.05, 0.1, 0.15, 0.2
\item \textbf{Possible convolutional filter sizes:} 3, 5, 7
\item \textbf{Possible number of filters:} 2, 16, 32
\item \textbf{Possible sizes of fully connected layers:} 10, 50, 100
\end{itemize}

The parameters for evolution were the following:
\begin{itemize}
\item \textbf{Proportion of parents:} 25\%
\item \textbf{Maximum number of generations:} 10
\item \textbf{Population per generation:} 8
\item \textbf{Maximum generation stall for early stop:} 2
\end{itemize}

Each individual in a generation is trained 4 times and the average classification accuracy is used to determine its fitness. The training consists of 50 epochs with early termination and a mini batch size of 20 samples.

\pagebreak[4]

\begin{table}
\caption{Classification accuracy of spectrogram-only networks}
\begin{tabular}{lp{0.33\linewidth}p{0.33\linewidth}} \hline
 category & size (median of top 3 in the bucket) & accuracy (median of top 3 in the bucket) \\ 
 \hline
 low accuracy & $0.3 \times 10^3$ & $50.1\%$ \\
 medium accuracy & $5.7 \times 10^6$ & $76.1\%$ \\
 high accuracy & $25.6 \times 10^6$ & $94.8\%$ \\ \hline
\end{tabular}
\label{tab:accuSpect}
\end{table}

\clearpage

\begin{table}
\caption{Classification accuracy of recording-condition only networks}
\begin{tabular}{lp{0.33\linewidth}p{0.33\linewidth}} \hline
 category & size (median of top 3 in the bucket) & accuracy (median of top 3 in the bucket) \\ \hline
 low accuracy & $0.4 \times 10^3$ & $51.3\%$ \\
 high accuracy & $0.6 \times 10^6$ & $61.7\%$ \\ \hline
\end{tabular}
\label{tab:accuCond}
\end{table}

\clearpage

\begin{table}
\caption{Improvement in ovenbird classification accuracy obtained by adding a recording-condition only network to a  spectrogram-only network.}
\begin{tabular}{p{0.20\linewidth}p{0.22\linewidth}p{0.22\linewidth}p{0.22\linewidth}} \hline
  & spectrogram network of low accuracy ($50.1\%$) & spectrogram network of medium accuracy ($75.6\%$) & spectrogram network of high accuracy ($94.77\%$) \\ \hline
 recording-condition network of low accuracy ($51.3\%$) & $+20.3\%$ (combined $70.3\%$) & $+0.2\%$ (combined $75.8\%$) & $+0.1\%$ (combined $94.8\%$) \\
 recording-condition network of high accuracy ($61.7\%$) & $+21.8\%$ (combined $71.8\%$) & $+0.4\%$ (combined $76.0\%$) & $+4.3\%$ (combined $99.0\%$)\\ \hline
\end{tabular}
\label{tab:accuCond}
\end{table}

\clearpage

\begin{table}
\centering
\caption{Classification accuracy of combined networks (uses median of top 3 in the category)}
\begin{tabular}{lcc} \hline
 category & size & accuracy \\ \hline
 low accuracy & $0.3 \times 10^6$ & $69.8\%$ \\
 medium accuracy & $1.5 \times 10^6$ & $77.1\%$ \\
 high accuracy & $25.6 \times 10^6$ & $99.1\%$ \\ \hline
\end{tabular}
\label{tab:accuComb}
\end{table}

\clearpage

\begin{table}
\caption{Confusion matrix of the classification accuracy of recording-condition only networks.}
\begin{tabular}{lcc} \hline
  & Ovenbird song present & Ovenbird not present \\ \hline
 network returns Ovenbird & $24.3\%$ & $4.6\%$ \\
 network returns non-Ovenbird & $75.7\%$ & $95.4\%$ \\ \hline
\end{tabular}
\label{tab:confCond}
\end{table}

\clearpage

\begin{table}
\caption{Confusion matrix of the classification accuracy of recording-condition only networks.}
\begin{tabular}{p{0.22\linewidth}p{0.16\linewidth}p{0.24\linewidth}p{0.24\linewidth}} \hline
  & Ovenbird present & Ovenbird song not present (Ovenbird habitat) & Ovenbird song not present (non-Ovenbird habitat) \\ \hline
 network returns Ovenbird & $24.3\%$ & $8.2\%$ & $1.0\%$\\
 network returns non-Ovenbird & $75.7\%$ & $91.8\%$ & $99.0\%$ \\ \hline
\end{tabular}
\label{tab:confCondBD}
\end{table}

\clearpage

\begin{table}
\caption{Confusion matrix of the classification accuracy of spectrogram-only networks.}
\begin{tabular}{lcc} \hline
  & Ovenbird song present & Ovenbird not present \\ \hline
 network returns Ovenbird & $97.7\%$ & $8.0\%$ \\
 network returns non-Ovenbird & $2.3\%$ & $92.0\%$ \\ \hline
\end{tabular}
\label{tab:confSpec}
\end{table}

\clearpage

\begin{table}
\caption{Confusion matrix of the classification accuracy of spectrogram-only networks.}
\begin{tabular}{p{0.22\linewidth}p{0.16\linewidth}p{0.24\linewidth}p{0.24\linewidth}} \hline
  & Ovenbird present & Ovenbird song not present (Ovenbird habitat) & Ovenbird song not present (non-Ovenbird habitat) \\ \hline
 network returns Ovenbird & $97.7\%$ & $6.0\%$ & $10.0\%$\\
 network returns non-Ovenbird & $2.3\%$ & $94\%$ & $90.0\%$ \\ \hline
\end{tabular}
\label{tab:confSpecBD}
\end{table}

\clearpage

\begin{table}
\caption{Confusion matrix of the classification accuracy of combined networks.}
\begin{tabular}{p{0.28\linewidth}p{0.24\linewidth}p{0.24\linewidth}} \hline
  & Ovenbird song present & Ovenbird not present \\ \hline
 network returns Ovenbird & $98.8\%$ ($+1.1\%$ from recording condition) & $0.6\%$ ($-7.4\%$ from recroding condition) \\
 network returns non-Ovenbird & $1.2\%$ ($-1.1\%$ from recording condition) & $92\%$ ($99.4\%$ from recroding condition) \\ \hline
\end{tabular}
\label{tab:confComb}
\end{table}

\clearpage

\begin{table}
\caption{Confusion matrix of the classification accuracy of combined networks.}
\begin{tabular}{p{0.22\linewidth}p{0.16\linewidth}p{0.24\linewidth}p{0.24\linewidth}} \hline
  & Ovenbird present & Ovenbird song not present (Ovenbird habitat) & Ovenbird song not present (non-Ovenbird habitat) \\ \hline
 network returns Ovenbird & $98.8\%$ ($+1.1\%$ from recording condition) & $0.6\%$ ($-5.4\%$ from recording condition) & $0.6\%$ ($-9.4\%$ from recording condition)\\
 network returns non-Ovenbird & $1.2\%$ ($-1.1\%$ from recording condition) & $99.4\%$ ($+5.4\%$ from recording condition) & $99.4\%$ ($+9.4\%$ from recording condition) \\ \hline
\end{tabular}
\label{tab:confCombBD}
\end{table}

\clearpage

\clearpage 

\begin{enumerate}
\item Figure 1: The general architecture of the combined networks. Each network (column) is evolved independently and then integrated for the final step.
\item Figure 2: Spectrogram of an Ovenbird call. Signal is in the center of the spectrogram. The image is composed of four spectrograms stacked on top of each other. The red lines were added to this Figure to show the separation of the four different spectrograms used; they were not visible to neural networks.
\item Figure 3: Markers show the accuracy of evolved networks and DenseNet47 against their size. The line plots show the median accuracy of the top three most accurate ANNs in each bucket for buckets with three or more NNs. Error bars show the difference between the median and the other top three networks for buckets that contain more than one network. 
\end{enumerate}

\end{document}